\documentclass[a4paper]{ar}
\usepackage[numbers]{natbib}
\usepackage{hyperref}
\usepackage[usenames,dvipsnames]{xcolor}
\usepackage{gensymb}
\usepackage{hyperref}
\hypersetup{
	bookmarks=true,
	unicode=false,
	pdftoolbar=true,
	pdfmenubar=true,
	pdffitwindow=false,
	pdfstartview={FitH},
	pdftitle={My title},
	pdfnewwindow=true,
	colorlinks=true,
	linkcolor=red,
	citecolor=blue,
	filecolor=magenta,
	urlcolor=cyan
}
\DeclareGraphicsExtensions{.eps,.pdf}
\newcommand{\etal}{{\it et al.}}

\begin{document}
\markboth{L.\,Wollmann}{Heusler 4.0}
\title{Heusler 4.0: Tunable Materials}

\author{Lukas~Wollmann,$^{1}$ Ajaya~K.~Nayak,$^{1,2}$ Stuart~S.~P.~Parkin,$^{2}$ and Claudia Felser$^{1}$
\affil{$^1$Max-Planck-Institut f{\"u}r Chemische Physik fester Stoffe, Dresden, Germany, D-01187}
\affil{$^2$Max-Planck-Institut f{\"u}r Mikrostruktur Physik, Halle, Germany, D-06120}}

\begin{abstract}
Heusler compounds are a large family of binary, ternary and quaternary compounds that exhibit a wide range of properties of both fundamental and potential technological interest. The extensive tunability of the Heusler compounds through chemical substitutions and structural motifs makes the family especially interesting. In this article we highlight recent major developments in the field of Heusler compounds and put these in the historical context. The evolution of the Heusler compounds can be described by four major periods of research.  In the latest period, Heusler 4.0 has led to the observation of a variety of properties derived from topology that includes: topological metals with Weyl and Dirac points; a variety of non-collinear spin textures including the very recent observation of skyrmions at room temperature; and giant anomalous Hall effects in antiferromagnetic Heuslers with triangular magnetic structures. Here we give a comprehensive overview of these major achievements and set research into Heusler materials within the context of recent emerging trends in condensed matter physics.
\end{abstract}

\maketitle
\tableofcontents
\clearpage

\clearpage
\section{HEUSLER 1.0 -- TURNING NON-MAGNETIC ELEMENTS INTO FERROMAGNETIC MATERIALS \label{part:heusler1}}
The history of Heusler compounds dates back to 1903, the year of Friedrich Heusler's seminal contribution~\cite{Heu1903,HSH1903} in the {\it Verhandlungen der deutschen physikalischen Gesellschaft}, where he announced the discovery of a ferromagnetic material at room temperature, that is surprisingly formed from the elements, Cu, Mn, and Al that show no magnetism at room temperature.  Later ferromagnetism was found in other compounds formed from Cu, and Mn but with several other elements ${Z} = \rm Sb,~Bi,~Sn$. Today, this perhaps does not seem so surprising especially since the concepts of antiferromagnetism and ferrimagnetism were introduced by Louis N{\'e}el in the 1930s-1940s~\cite{Neel1936, Neel1953}.  However, these phenomena were unknown in 1903, which made Heusler’s work a major finding. The structure of the compound that Heusler prepared was also unknown in 1903, although Heusler realized that a chemical compound must have been formed. He thus anticipated what is today widely understood and accepted: Heusler compounds form a special class of materials, that are located at the border between compounds and alloys, and which combine features of both, namely, the chemical stability of a covalent lattice from which the Heusler compound is constructed, while single sites within the lattice can be substituted by different species and thereby behave as single-site alloys. In a nutshell, covalency and tunability best describe the uniqueness of this materials class. It was not until 1934 that Otto Heusler, Heusler's son~\cite{HeuslerO1934}, and Bradley~\cite{BR1934}, determined the crystal structure of Cu$_2$MnAl. Otto Heusler noted the possibility of another type of crystalline order that is nowadays termed an \textsl{invererse/inverted} Heusler compound with the spacegroup $T_2^d$~\cite{HeuslerO1934} in the notation of Schoenflies or $216$ in today’s space group classification. The Heusler structure can be described as intertwined cubic and rocksalt lattices, or as four interpenetrating \textit{fcc} sublattices, of which two are formed from the same element.

\begin{figure}[ht]
	\includegraphics[width=1\linewidth]{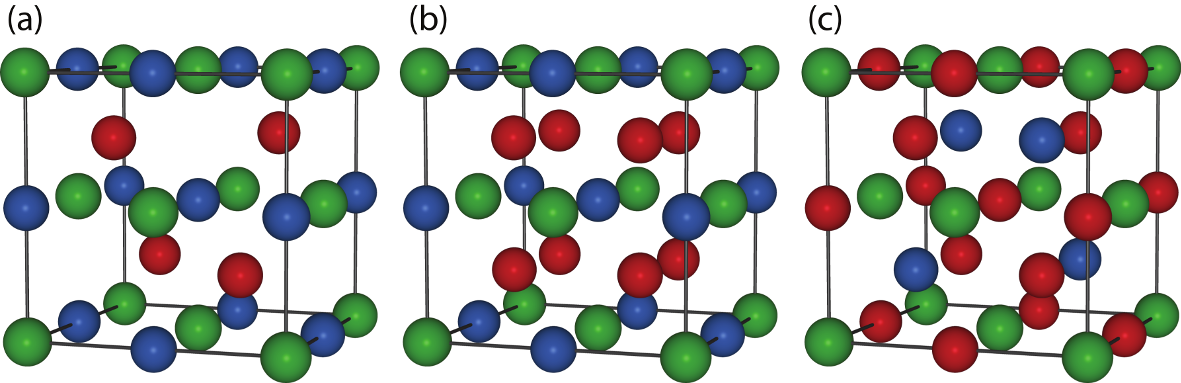}
	\caption{\label{fig1}(Color online)~The three prototypical Heusler structures, wherein $X$ atoms are represented by red spheres, $Y$ atoms by blue/light blue and $Z$ atoms by green spheres:~(\textit{a})~\textit{half}-Heusler,~(\textit{b})~\textit{regular} Heusler,~(\textit{c})~\textit{inverse} Heusler structure.}
\end{figure}
Shortly after Heusler’s discovery, Nowotny and Juza, published results on a different group of materials, all main-group element compounds, namely LiMgAs (Nowotny) and CuMgAs (Juza), that  are nowadays referred to as Nowotny-Juza-Phases. The connection between the Nowotny-Juza phases and the Heusler compounds, was established by L. Castelliz, who first synthesized NiMnSb~\cite{Castelliz1951,Castelliz1952}, as part of the compositional series Ni$_{2-x}$MnSb ($0\leq x \leq 1$). NiMnSb and the Nowotny-Juza-Phases are now described as \textit{half}-Heusler compounds in which one of the 4 \textit{fcc} sub-lattices of the \textit{full} Heusler is empty.  By filling this 4$^{\rm th}$ sub-lattice a series of compounds can be formed between \textit{half} and \textit{full} Heuslers, which we can describe as $XYZ$ and $X_2YZ$, respectively, where $X$, $Y$ are transition metal elements and $Z$ is a main group element. The \textit{full} Heusler compounds have several variants including the \textit{inverse} structure in which one of the $X$ elements is swapped with $Y$, and quaternary Heuslers in which one of the $X$ is replaced by a 4$^{\rm th}$ distinct element.  

\begin{table}[h]
\centering
  \begin{tabular}{ l | l | ccccc}
Type 	& Order 		&  4$d^{\rm HC}$& 4$c$	& 	4$b$& 	4$a$ & Example	\\\hline
Fluorite	&C1 			&  {\it A}  & 		& 	{\it E}		&	{\it E}		& Ca$^{\rm HC}$F$_2$\\
Juza 	& C1$_{\rm b}$ 	&  {\it E}  & 		&	{\it A}		&	{\it A$'$} &LiMgN$^{\rm HC}$	\\
Nowotny & C1$_{\rm b}$ 	&  {\it E}	&		&	{\it T}	&{\it A} &	MgAgAs$^{\rm HC}$	\\\hline
Nowotny & C1$_{\rm b}$ 	&  {\it T}	&		&{\it A}		&{\it E} 	& MgCu$^{\rm HC}$Sb	\\
{\it Half} -- Heusler	& C1$_{\rm b}$	&  {\it T}	&	&{\it T}$'$&{\it E} & MnCu$^{\rm HC}$Sb	\\
{\it Full} -- Heusler	&  L2$_{\rm 1}$	&  {\it T}	&	{\it T}&{\it T}$'$&{\it E}& Cu$_2$$^{\rm HC}$MnAl \\
{\it Inverse} -- Heusler& X$_{\rm a}$	&  {\it A}	&	{\it T}	&	{\it A}	&	{\it E}& Li$_2$AgSb \\ 
  \end{tabular}
\vspace*{8mm}\caption{The different site occupations as they are found in cubic Heusler and Heusler-related--structures. See also ~\cite{GFP11}.}
\label{tab1}
\end{table}

Thus, it is clear that the Heusler name now refers to a broad and extensive family of compounds. Furthemore, all of these variants can be subjected to various structural distortions including a tetragonal elongation or compression along one of the cubic crystal axes, or a distortion along the [111] direction that leads to an hexagonal structure~\cite{GFP11}. Finally, super-structures are sometimes found that can arise from chemical ordering in non-stoichiometric compounds or from modulation of the structure and structural phase transitions, for example, in the case of shape memory alloys. The key differences between these various Heusler compounds are highlighted in Table~\ref{tab1}.

The three main prototypical Heusler stoichiometric, chemically ordered structural types, namely the \textit{half}, \textit{regular} and \textit{inverse} are shown in Figure\,\ref{fig1}.\\
Early reports on Heusler phases~\cite{Webster1971} appear in the context of $X_2$Mn$Z$ alloys, where $X=\rm Ni,~Pd,~Au$ and $Z=\rm Al,~Si,~Ga,~Ge,~In,~Sn,~Sb$, with a particular focus on their chemical and magnetic order, and an ordering transition that can be induced by composition or  by temperature, or pressure.~\cite{Castelliz1951,WT+1967,MPW1959}. Remarkably, a ferro- to antiferromagnetic transition was observed in Pd$_2$MnIn$_{1-x}$Sn$_{x}$~\cite{WR1977}.  Webster and co-workers were the first to explore the magnetic properties of the Co$_2$Mn$Z$ compounds, which they stated were quite different from previously studied Heuslers, as they incorporate elements other than Mn that carry a substantial local magnetic moment~\cite{Webster1971}. They did not anticipate the great interest of the  scientific community in these materials that has occurred over the past decade.\\ It is worth mentioning that cousins of the original Heusler-type were found in the 1960--1970s at the IBM Research Laboratory, San Jose (CA). Jim Suits explored the Rh$_2$-based Heusler compounds, and found a structural transformation at temperatures of about 700--800~K, with a very large deformation amplitude $\varepsilon = c_{\rm tet}(c_{cub} - 1)$ of about 17\% which is of the same size as the Mn$^{3+}$ Jahn-Teller-ion containing spinels~\cite{Suits1976a,Suits1976}. Some of these compounds were non-magnetic and some antiferromagnetic, and some showed tetragonally distorted derivatives which have become of considerable interest today.\\
It was not until much later, that the Co$_2$-based Heusler compounds received renewed attention in the context of half-metallic ferromagnetism (HMF). Half-metallic ferromagnets, exhibit metallic behavior in one spin channel and an insulating behavior in the other, and thus are of great interest because they intrinsically should have fully spin polarized electronic states at the Fermi energy~\cite{FFB2007}.\\
The first example of the Heusler HMF was first realized by de Groot \etal~for the case of NiMnSb~\cite{GME+1983}. Such materials are also of great interest for spintronic applications, for example, as magnetic electrodes in magnetic tunnel junctions.  Magnetic tunnel junction are composed of two magnetic electrodes separated by a thin tunnel barrier. The current crossing the tunnel barrier from one electrode to the other depends on the magnetic configuration of the electrodes. When the momets of the two electrodes are aligned parallel to one another the current can flow easily but when the moments of the two electrodes are aligned exactly anti-parallel to one another the tunneling current will be reduced, and, for the case of HMF, should go to zero. \\
While semiconducting Heusler compounds are usually not magnetic, magnetism can be introduced by the introduction of $RE$ elements, $RE$PtBi or Mn on the 4$b$ position.  Whilst the Mn containing compounds become metallic, the $REYZ$ remain semi-conducting due to the localization of the 4$f^n$-electrons. The semiconducting behavior of Mn$YZ$ (commonly written as $Y$Mn$Z$) is conserved only in one spin-channel , and therefore these compounds have been coined "half-metallic ferromagnets", in which one spin-channel exhibits semi-conducting or insulating characteristics caused by a minority gap, while the majority spin-channel is metallic. NiMnSb is of this type and has 22 valence electrons. This compound has four excess electrons as compared to the semi-conducting Heuslers that have a valence electron count of $N_{\rm V}=18$. Interestingly, these excess electrons fill one spin-channel and thereby form a spin magnetic moment of $M_{spin} = 4\,\mu_{\rm B}$.  In this way these compounds follow almost exactly the  Slater-Pauling rule whereby $M_{spin} = N_{\rm V} - 18$. We will return to this point later (Sec.\,\ref{part:heusler2}.\ref{sec:mnhmf}). The early work on NiMnSb emphasized its  magneto-optical properties that was stimulated by the large magneto-optical Kerr angle of 1.27$\degree$ found in PtMnSb, one of the other early HMFs~\cite{GME+1983}. 

\clearpage
\section{HEUSLER 2.0 -- HALF METALLICITY\label{part:heusler2}}
\begin{figure}[ht]
	\includegraphics[width=\textwidth]{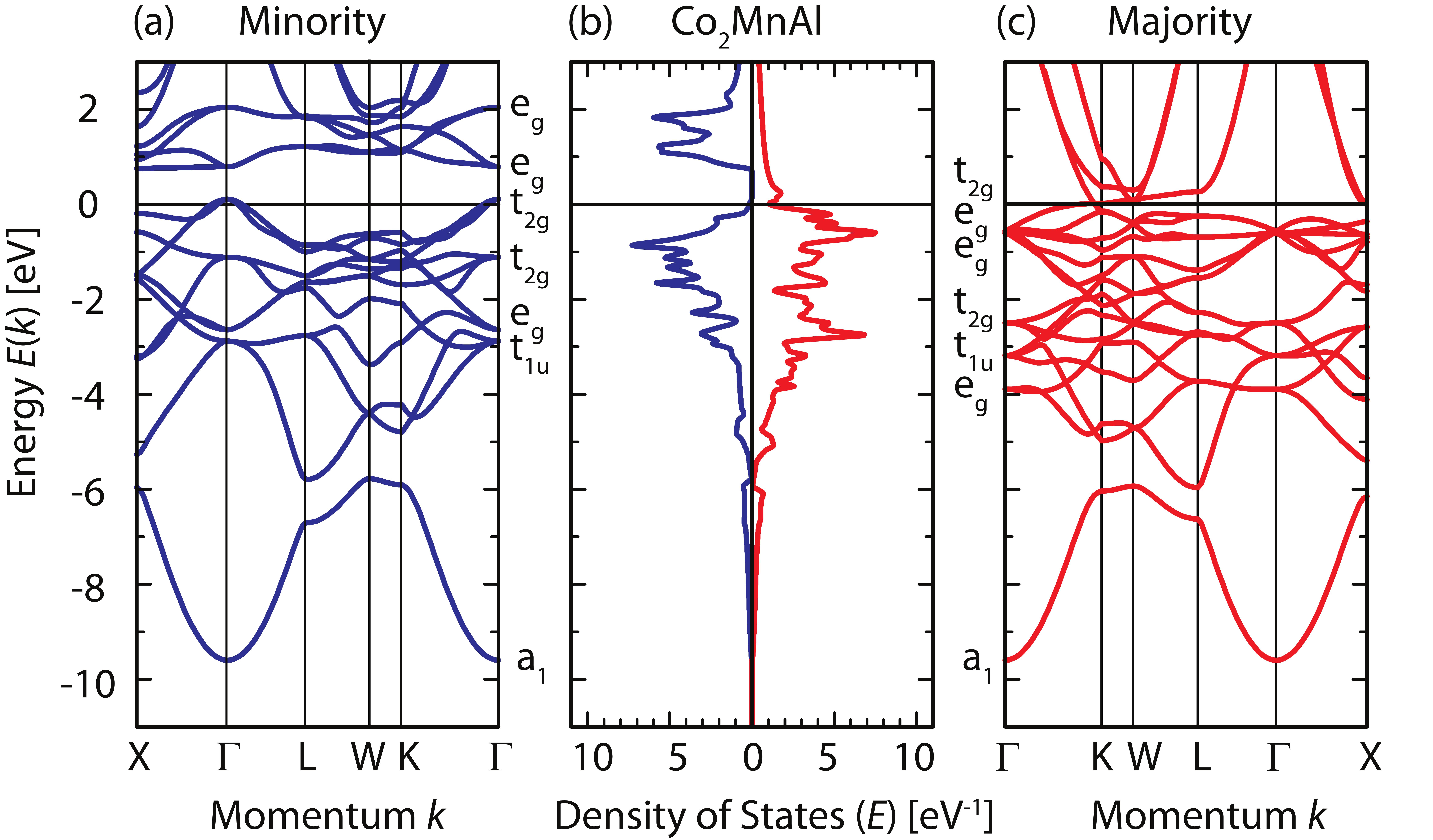}
	\caption{\label{fig2}(Color online) Band structure of Co$_2$MnAl depicting the E($k$) dispersion relation for the minority~(\textit{a}) and majority~(\textit{c}) spin channels, together with the density of states~(\textit{b})}
\end{figure}

\subsection{Co-based Half-metallic Ferromagnets {\label{sec:cohmf}}}
It was Ishida \etal, who first began exploring HMF in Co$_2$-based Heusler compounds, such as  Co$_2$MnSn, Co$_2$TiAl, Co$_2$TiSn, using LDA methods to calculate their electronic structure.  However, Ishida \etal~did not obtain HMF in these compounds~\cite{IAK+1982} because the LDA~\cite{VWN1980} formalism did not allow for a minority gap to open.  This was later confirmed by K\"ubler \etal~\cite{KWS1983}, who noted, however, that {\it peculiar transport properties} are to be expected for these nearly HMF-type compounds. It was in the same year, 1983, that K{\"u}bler studied the formation and coupling of the magnetic moments in Co$_2$MnSn and similar Co-based alloys, and remarked how their spin magnetic moments follow a linear behavior with the number of valence electrons. K{\"u}bler \etal~quickly recognized the peculiar role of Mn in the $X_2$Mn$Z$ type compounds. They could reproduce the ferro- to antiferromagnetic transition from Pd$_2$MnIn to Pd$_2$MnSn and elaborated on the formation of localized magnetic moments in purely itinerant magnets, triggered by the strong exchange splitting of the $d$-states attributed to the Mn-site. These findings were soon put into context, and chemical trends were established of which the backbone is the Slater-Pauling (SP) rule. The Slater--Pauling rule was first expressed as a function of the magnetic valence, but soon reformulated in a more general manner. Originally the SP-rule covered elements and binary alloys~\cite{Friedel1958}, but the extension to ternary alloys led to the wide success of the SP-rule as a very useful tool for assessing the expected magnetic moment of a ternary HMF. The Slater-Pauling rule describes the interplay of electron filling and the resulting magnetic moment. In Co-based \textit{full}-Heusler alloys, a valence electron count of 24 leads to a zero net moment as both spin-channels are occupied by 12 valence electrons each. Filling or depleting electrons leads to a net integer spin-moment $M=N_{\rm V}-24$.\\
Further contributions to the theory of Heusler alloys showed how the SP-rule could be described using molecular orbital coupling schemes and symmetry analysis. Fecher \etal~emphasized that an upper bound of approximately $6\,\mu_{\rm B}$~\cite{OFB+2011} is unlikely to be exceeded, as the $s$-states which were to be populated in the majority spin-channel are strongly dispersed, which would lead to an energetically unfavorable situation. Furthermore, it could be shown that some compounds involving late transition metals $Y=\rm Cu,~Zn$ , e.g. Mn$_2$ZnSi, do not obey the SP-rule as $M=N_{\rm V}-24$, but another situtation occurs in which doubly degenerate $d$-states are shifted below a set of triply degenerate states caused by the symmetry of the crystal structure (X$_{\rm a}$, no inversion), such that the equation $M=N_{\rm V} - 28$ guides the emerging magnetic moments~\cite{SOS+2013}.\\
The Slater-Pauling rule is a simple yet powerful tool for the prediction of half-metallic ferromagnets. In principle, one would not expect a simple relationship between a compound's magnetic moment, and the critical magnetic ordering temperature, that depends sensitively on the electronic structure. In the Heisenberg picture, this depends on an atom's nearest neighbors. In Co-based Heusler compounds, a simple relationship wherein $T_C \propto M$ has been found~\cite{KFF2007a}. This is another beautiful example of the fundamental tunability of Heusler compounds, limited only by their chemical stability. The evolution of the Curie temperature is traced back to two competing factors in terms of the spherical approximation (SPA), namely the decreasing average exchange energy counterbalanced by an increasing total moment with increasing number of valence electrons. It was concluded that the double-/kinetic exchange mechanism of Zener provides the key concept to the understanding of the observed behavior~\cite{KFF2007a,Kubler09}.\\
The beginning of a phase of intense research on the Heusler class of compounds that lay the foundation for  the current perspective of Heusler compounds as multi-potential materials, possibly providing answers for many materials science challenges, was in the beginning of the 1980s. The use of half-metallic ferromagnets in spin-valve structures should lead to out-of-scale magnetoresistance, from which the field of spintronics received an activity boost. Spin-valve structures based on NiMnSb did not appear to be promising in the beginning so mechanisms leading to a possible suppression of the expected high TMR values were explored. By means of relativistic {\it first principles} calculations it was shown that spin-orbit coupling induces a finite but negligibly DOS for 3$d$-based Heusler compounds, while this effect is more severe for PdMnSb and PtMnSb. After these less successful attempts, the incorporation of Co$_2$-based Heusler alloys, such as Co$_2$Cr$_{0.6}$Fe$_{0.4}$Al~\cite{BFJ+03} and Co$_2$MnSi into spin-valve structures succeeded in delivering the expected MR-values in the range of $\rm MR\approx 2000~\%$, after many years of materials and interface engineering~\cite{LHT+2012}. Quite recently, in 2014, spin-polarized photoemmission experiments conducted by Jourdan \etal, supported by theoretical calculations, provided proof of half-metallic ferromagnetism in Co$_2$-based Heusler alloys (in the exemplary compound Co$_2$MnSi) (within the experimental accuracy)~\cite{JMB+2014}.

\subsection{Mn-based Half-metallic Ferrimagnets{\label{sec:mnhmf}}}
In addition to NiMnSb and Co$_2$MnSi, Mn$_2$VAl has also attracted a lot of attention, as it was early identified as a HMF by numerical methods~\cite{WP1999}. Nevertheless, experimental and theoretical studies have largely focused on the Co-based compounds. Renewed interest in the Mn-based compounds was triggered by the discovery of structurally distorted cousins of the cubic systems: namely, the tetragonally distorted Heusler compounds. The most renowned member, Mn$_3$Ga~\cite{KK1970}, was already studied in the 1970s, yet the potential for spintronic applications was not recognized until the late 2000s~\cite{BFW+2007,WBF+2008}.\\
The first example of a member of the Mn$_2$- family is without doubt Mn$_2$VAl. It was synthesized by Kopp~\cite{KW1969} and Nakamichi~\cite{NI1978}, but misinterpreted as a ferromagnet, which was later corrected by neutron diffraction measurements~\cite{INY+1983}, before it was studied within the scope of half-metallicity~\cite{WP1999}. Mn-based materials received renewed attention in the late 2000s, when ferrimagnetic Heusler compounds where proposed as free-magnetic layers for spin-valve structures, such as, for example, magnetic tunnel junctions, in which a spin-polarized current is used to trigger the switching of the free-layer. For a long time Mn$_2$VAl was the only known member of that class until low moment magnets for spintronics/electrodes/memory received renewed attention, which led to the discovery of ferrimagnetic Heuslers with low magnetization. Further half-metallic ferrimagnetic Mn$_2$V$Z$-Heusler compounds were predicted by DFT based computations~\cite{GDP2002,OeGS+2006,WCK+2014}. With these compounds it was realized that Slater-Pauling behavior is possible even when the valence electron count is less than 24.  Then negative moments are formally computed, which reflects the relative magnitudes of the individual sub-lattice moments, when the orientation of these sub-lattices is fixed with respect to the crystal lattice. This can lead to a compensation point, where the two sub-lattice moments are identical. When the composition is systematically varied within a given compositional series, this is particularly interesting, as the a sign change of the net magnetic moment may also give rise to a sign change in other properties.\\
The formation of magnetic moments on the Mn atoms at site 4$b$ was explained by K{\"u}bler~\cite{KWS1983}, and is found to be correct even within the Mn$_2$-Heusler family. The Mn atoms at the $X\mathopen{}\left(4c/4d\right)\mathclose{}$ site are found to have small moments comparable to that of the Co atoms at the same positions within the Co$_2$-Heusler class.
A high moment, of about 3\,-\,4$\,\mu_{\rm B}$ at site $Y\mathopen{}\left(4b\right)\mathclose{}$~\cite{WCK+2014} was observed from the very beginning of studies on these compounds, and was identified as a characteristic feature of the Mn-$Z$ layer. The Mn$_2$-Heuslers can also form another type of order that is today termed as {\it inverted}. It emphasizes the fact, that an atomic swap/exchange is observed in comparison to the prototypical Cu$_2$MnAl type, within which the $X$ atoms evenly occupy the site 8$c$ in the inversion symmetric spacegroup 225.
Independent of these types of order, the Slater-Pauling rule remains valid even for the inverted type.

\begin{figure}[ht]
\centering
\includegraphics[width=.7\textwidth]{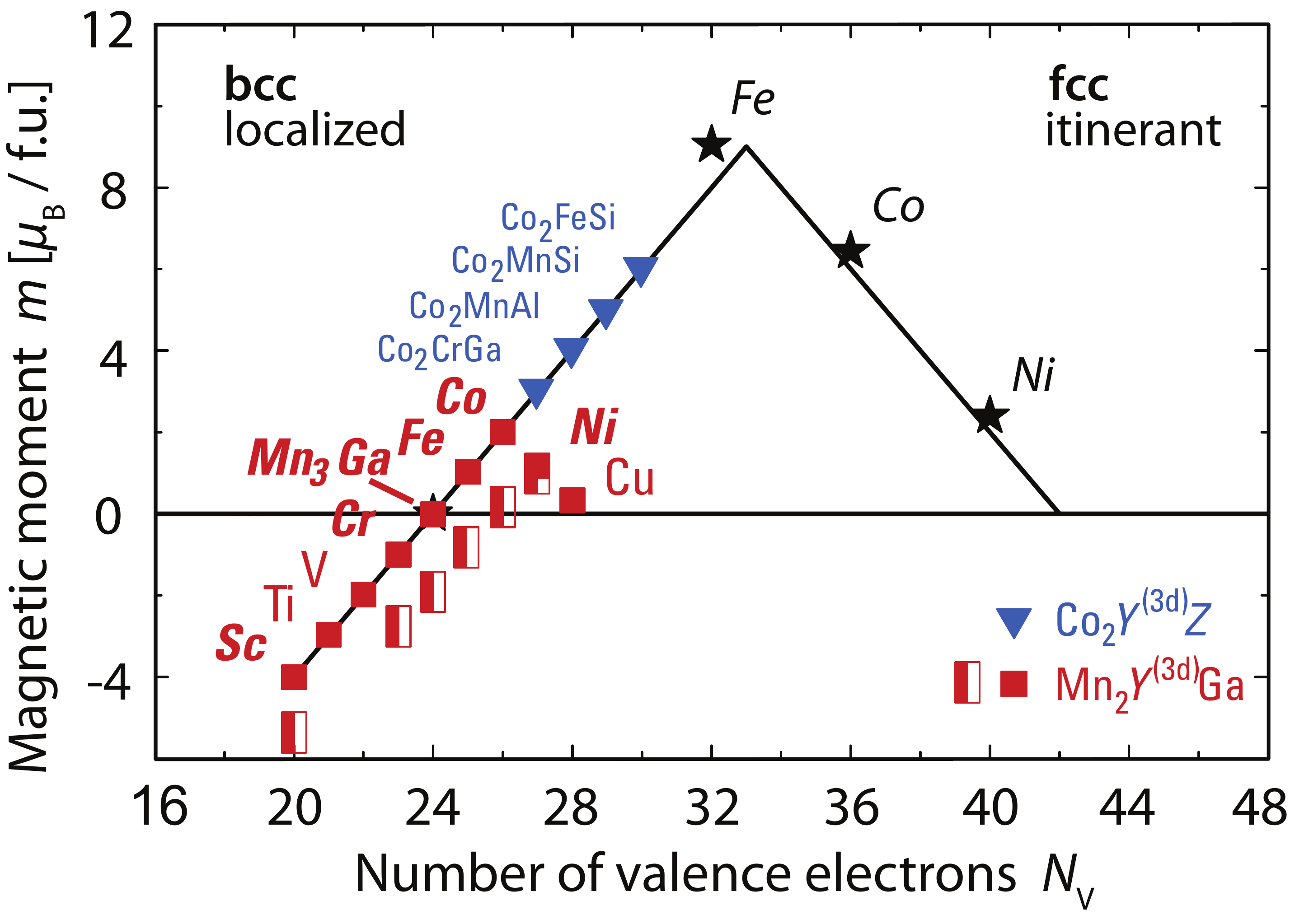}
\caption{\label{fig3}(Color online) Schematic representation of the Slater-Pauling behavior for a set of selected Mn$_2Y$Ga and Co$_2YZ$ compounds. Based on Refs.~\cite{WCK+2014,WCK+2015}}
\end{figure}

Further ferrimagnetic and possibly half-metallic systems have been studied for their magnetic ground-states, while these studies often refer to simple numerical calculations, neglecting other types of magnetic and structural order.
Mn$_3$Ga is a material that exhibits features of \textit{regular} $X_2YZ$ and \textit{inverse} $XYXZ$ ordered Heusler compounds at the same time. At first, the even distribution of Mn atoms at the 8$c$ position, that resembles \textit{regular} order in $X_2YZ$, while the occupation of Mn atoms 
of the 4$b$ position is indeed a feature of \textit{inverse} order in the Mn$_2$-Heusler class, as for instance in Mn$_2$CoAl, where Mn and Co are located at 4$d$/4$c$ and Mn and Al at 4$b$ and 4$a$. In theory Mn$_3$Ga crystallizes in the centrosymmetric space group 225 (prototype BiF$_3$) showing compensated ferrimagnetism~\cite{WKF+2006}.
That said, it is found in various studies, that systems incorporating early transition metals as $Y$ elements prefer the \textit{regular} order, while choosing late transition metals the \textit{inverse} order is preferred~\cite{BLB1974,LZM+2008,WCK+2014}, where the early and late transition metals are separated by Mn. This observation is known as {\it Burch's rule}~\cite{BLB1974}.
In experimental studies bulk-Mn$_3$Ga has been found to stabilize in a tetragonal lattice (see Sec.\,\ref{part:heusler4}.\ref{sec:stt} and \ref{part:heusler4}.\ref{sec:tetra-cchmf})

\subsection{Compensated Ferrimagnets{\label{sec:ccfim}}}
\begin{figure}[ht]
	\includegraphics[width=.95\textwidth]{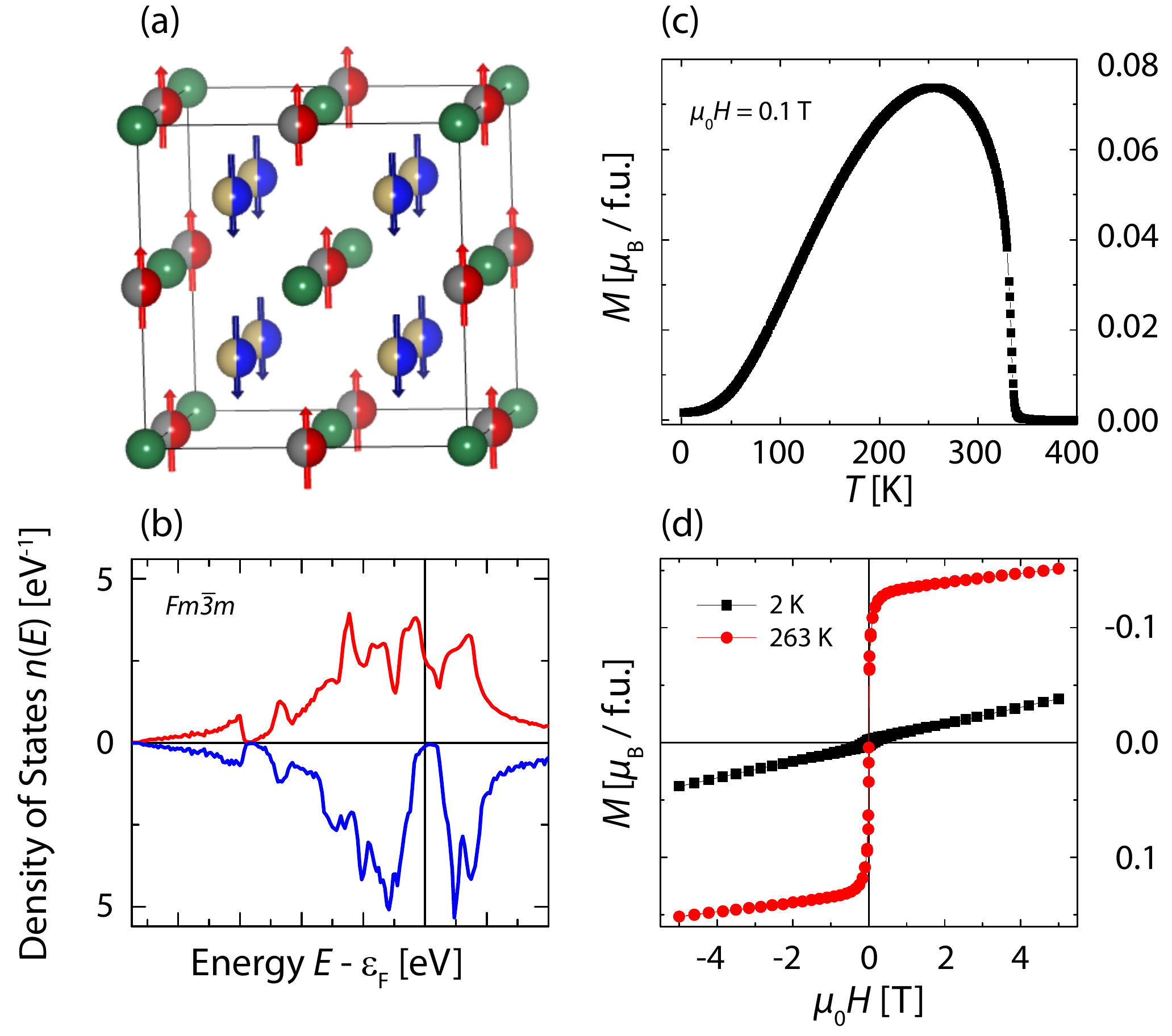}
	\caption{\label{fig2-4}(Color online)~(\textit{a})~Crystal structure of the L2$_{\rm 1}$-type cubic Heusler compound Mn$_{1.5}$FeV$_{0.5}$Al (space group $Fm\bar{3}m$) and the orientation of the magnetic moments on different sub-lattices. Different atoms are represented by balls with different colors.~(\textit{b})~The spin resolved density of states for Mn$_{1.5}$FeV$_{0.5}$Al.~(\textit{c})~Temperature dependence of magnetization ($M\left(T\right)$) measured in a field of $0.01\,\rm T$ for Mn$_{1.55}$V$_{0.3}$Fe$_{1.08}$Al$_{0.7}$.~(\textit{d})~Field dependence of magnetization ($M\mathopen{}\left(H\right)\mathclose{}$) measured at $2\,\rm K$ and $263\,\rm K$ for Mn$_{1.55}$V$_{0.3}$Fe$_{1.08}$Al$_{0.7}$.}
\end{figure}
Despite their potential for obtaining 100\% spin polarization, HMFs typically produce large magnetic dipole fields that can hinder the performance of spintronic devices that contain them. Therefore, materials that display 100\% spin polarization but with very low or, even more interestingly, zero net magnetic moment are of special interest, both technologically as well as scientifically. According to the Slater--Pauling~\cite{Slater1936,Pauling38} rule, which, as discussed above, also describes the HMF Heuslers, the total magnetic moment ($M$) of the L2$_{\rm 1}$ cubic Heusler compounds is given by $M_{spin} = N_{\rm V} −- 24$, where $N_{\rm V}$ is the number of valence electrons. Thus, Heusler compounds with 24 valence electrons should exhibit a zero net magnetic moment. The search for compensated ferrimagnetic Heuslers with 24 valence electrons has been focussed on the Mn-based Heusler compounds since the Mn atoms sit in an octahedral environment that results in a strongly localized magnetic moment. An example of a 24 valence electron based Heusler compound is Mn$_3$Ga, which is predicted to display half-metallicity in the cubic L2$_{\rm 1}$ structure. Unfortunately, cubic Mn$_3$Ga does not exist in the bulk form due to a tetragonal distortion that destroys both the compensated magnetic state and the half-metallic behavior. Recently, it has been shown that it is possible to achieve a compensated magnetic state in cubic thin films of Mn$_2$Ru$_x$Ga~\cite{KRS+2014,BTL+2015}. In this case, the compensated magnetic state was obtained in off-stoichiometric thin films having 21 valence electrons.\\

Recently, Stinshoff \etal~have found that the cubic L2$_{\rm 1}$ compound Mn$_{1.5}$FeV$_{0.5}$Al with 24 valence electrons exhibits a fully compensated ferrimagnetic state~\cite{SFF2016}. As shown in Figure\,\ref{fig2-4}\,(\textit{a}),  Mn$_{1.5}$FeV$_{0.5}$Al crystallizes in the \textit{regular} L2$_{\rm 1}$ cubic Heusler structure with space group $Fm\bar{3}m$. Al atoms (green spheres) occupy the 4\textit{a} position, V (grey) and MnI (red) atoms equally occupy the 4\textit{b} position, whereas, the 8\textit{c} position is occupied equally by MnII (green) and Fe (light yellow) atoms. A calculation of the site specific magnetic moment yields a larger localized moment of about $2.57\,\mu_{\rm B}$ for the Mn atoms that are sitting in the octahedrally coordinated 4b position, as compared with a smaller moment of $1.25\,\mu_{\rm B}$ for the tetrahedrally coordinated Mn atoms that sit in the 8c position. The V and Fe atoms exhibit much smaller moments of only $0.43\,\mu_{\rm B}$ and $0.23\,\mu_{\rm B}$, respectively. In this configuration a total moment of $3\,\mu_{\rm B}$ for Mn at the 4\textit{b} site and $2.96\,\mu_{\rm B}$ for Mn at the 8\textit{c} site is calculated. A nearly zero net magnetic moment is expected due to the antiparallel alignment of the moments on the two different sub-lattices. The calculation of the total density of states indicates the presence of a pseudo-gap in one of the spin directions, demonstrating a half-metallic character of Mn$_{1.5}$FeV$_{0.5}$Al (Figure\,\ref{fig2-4}). The experimental verification of the compensated magnetic structure was obtained from the $M\left(T\right)$ measurements shown in Figure\,\ref{fig2-4}\,(\textit{c}). The $M\left(T\right)$ measurement performed in a field of $0.01\,\rm T$ exhibits a Curie temperature ($T_{\rm C}$) of about $340\,\rm K$. As the temperature is reduced the magnetization drops to a nearly zero value at $2\,\rm K$, indicating the presence of a fully compensated magnetic state near zero temperature.  The $M\mathopen{}\left(H\right)\mathclose{}$ loop measured at $2\,\rm K$ exhibit a linear behavior with a nearly zero spontaneous magnetization. This confirms the completely compensated ferrimagnetic nature of Mn$_{1.5}$FeV$_{0.5}$Al.  The $M\mathopen{}\left(H\right)\mathclose{}$ hysteresis curve measured at $263\,\rm K$ displays a small residual moment of about $0.1\,\mu_{\rm B}$, mostly originating from a small compensation arising from the different temperature dependences of the sub-lattice magnetic moments.

\subsection{Spin-gapless Semiconductors {\label{sec:sgs}}}
Within the subset of Mn-based Heusler compounds, a particularly interesting material is Mn$_2$CoAl, which is the  first example of a spin-gapless semiconductor (SGS) in the Heusler family of compounds~\cite{OFF+2013}. Spin gapless Semiconductors were originally proposed by Wang in 2008~\cite{Wang2008} and are formed when a gapless semi-conductor is doped with magnetic ions. The \textit{fcc}-type band structure, that is an inherent feature of Heusler compounds, allows for this peculiar electronic feature. When there are between 18 and 30 valence electrons, several energy windows in the band structure have only weakly dispersed bands. A simplified molecular orbital diagram shows how a sequence of doubly and triply degenerate states are successively filled for 18, 21, 24 and 26 valence electrons. Mn$_2$CoAl with $N_{\rm V} =26$, naturally fulfills this requirement. {\"O}zdo\u{g}an \etal, have studied a set of quaternary LiMgPdSn-type materials, predicting five SGSs: CoFeCrAl, CoMnCrSi, CoFeVSi and FeMnCrSb, whereas FeTiVSi is almost a SGS, as the Fermi energy touches the edges of the conduction and valence bands~\cite{OSG2013}. Synthesized and characterized by Ouardi~\etal, Mn$_2$CoAl turned out to show a peculiar electronic structure, that was termed a {\it spin-gapless semiconducting state}~\cite{Wang2008,OFF+2013}, which depicts the band-gap in the minority spin-channel, accompanied by an indirect zero-band-gap in the majority spin-channel. The Curie temperature was found to be $T_{\rm C} = 720\,\rm K$, with a magnetic moment of about $2\,\mu_{\rm B}$. The carrier concentration below $300\,\rm K$ is nearly independent of temperature, while the Seebeck coefficient is vanishingly small. SGSs are expected to find applications in spintronic devices, especially semiconductor spintronics as the electronic excitations in the gapless state do not require a threshold energy, but the carriers, whether holes or electrons, remain completely spin-polarized. These materials thus may serve, for example, as spin-injectors.

\begin{table}[h]
	\centering
	\begin{tabular}{llllll}
		$N_{\rm V}$ 	& \multicolumn{5}{c}{Materials}\\\hline
		21 & FeVTiSi & CoVScSi & FeCrScSi & FeMnScAl& \\
		26 & Mn$_2$CoAl & CoFeCrAl & CoMnCrSi & CoFeVSi& FeMnCrSb \\
		28 &CoFeMnSi&&&&\\ 

	\end{tabular}
	\vspace*{8mm}\caption{Some examplary spin-gapless semiconductors, exclusively crystallizing in spacegroup 216.}
\end{table}

\clearpage

\section{HEUSLER 3.0 -- UNIAXIAL HEUSLER COMPOUNDS AND NON-COLLINEAR SPIN STRUCTURES \label{part:heusler4}}
\subsection{Tetragonal Heusler Compounds for Spin-Transfer Torque{\label{sec:stt}}}
\begin{figure}[ht]
	\includegraphics[width=.9\textwidth]{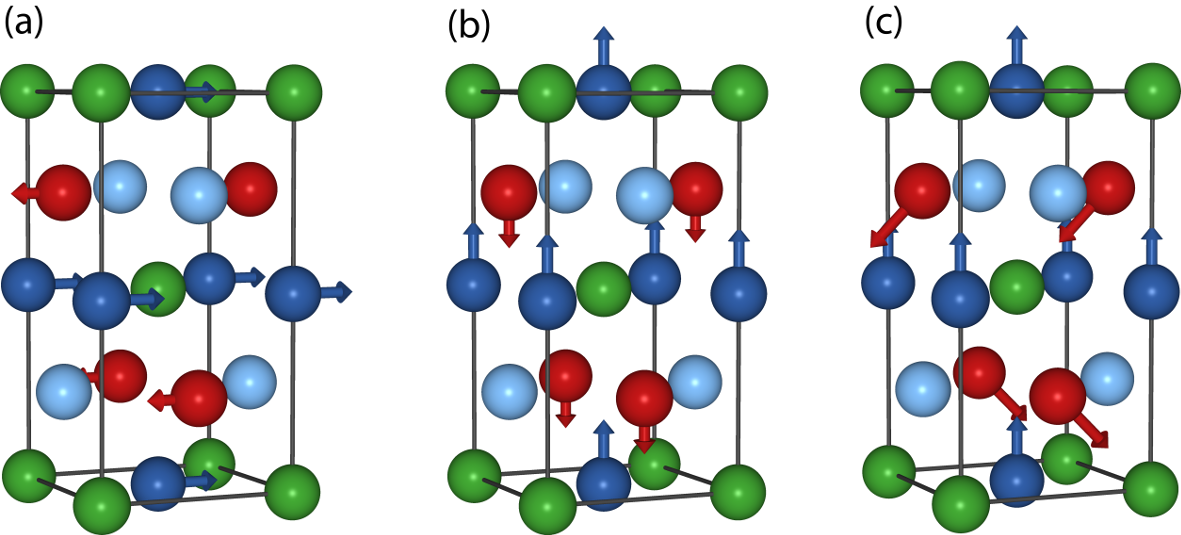}
	\caption{\label{fig5}(Color online) A set of generalized tetragonal Heusler structures (spacegroup 119) are shown depicting the four Wyckoff-Positions, that are to be occupied according to Mn (red, blue), $Y$ (light blue), Ga (green, $Z$) to achieve the \textit{inverse}-type order;~(\textit{a})~In-plane ($K_{\rm U} < 0$) and~(\textit{b})~out-of-plane ($K_{\rm U} > 0$) orientation of the magnetization.~(\textit{c})~ Exemplary non-collinear order for competing in-plane and out-of-plane anisotropy contributions.}
\end{figure}
Besides cubic Heusler compounds, which  show lots of interesting physics, their structural cousins, the tetragonal Heuslers, are widely studied today. The tetragonal structural modification lowers the symmetry, and unseen effects emerge. Tetragonal Heusler compounds were described by Suits in the 1970s~\cite{Suits1976,Suits1976a}, whereas major research began only in the 1990s when research focused on reversible structural phase transitions from cubic to tetragonal lattices, i.e. from austenite to martensite. Shape memory materials rely on this transition, of which Ni$_2$MnGa~\cite{UHK+96} and Mn$_2$NiGa~\cite{LDY+06} are the flagship materials. The latest boost in tetragonal Heusler compounds research began with the realization that the large magnetocrystalline anisotropy of Mn$_3$Ga~\cite{KK1970,BFW+2007} by Balke \etal, was of use for spintronic applications. While Co-based materials tend to show soft magnetic properties, the tetragonal distortion induces a preferred orientation of the magnetization towards the in-plane or out-of-plane directions (Figure~\ref{fig5}). Perpendicular magnetic anisotropy, with the magnetization pointing perpendicular to the film surface, is desired for high density memory and storage devices to overcome the current limitations, and to guarantee thermal stability. Magnetic memory bits can be switched by either magnetic field or via spin-polarized currents through the concept of spin-transfer-torque. Spin-transfer-torque (STT) refers to a torque that is exerted perpendicular to the magnetization, leading to a precessing magnetic moment, that finally switches to the opposite direction. The Slonczewski-Berger equation~\cite{Slonczewski96,Berger96} describes the dependence of the switching current density on materials properties such as magnetic moment $M$, anisotropy constant $K_{\rm U}$ and Gilbert damping parameter $\alpha$. As the STT technology requires materials with small switching currents while guaranteeing data retention/thermal stability, the Mn-based Heusler compounds have been explored in searches for new tetragonal phases.
That said, Alijani \etal~\cite{AWF+11} have studied the Mn-Co-Ga system~\cite{OKF+2012}, testing the stability of the tetragonal Mn$_3$Ga phase with the substitution  of Mn by Co: Mn$_{3-x}$Co$_x$Ga. A tetragonal phase is maintained until a critical Co concentration of $x_{\rm Co} = 0.5$. The subsequent members with higher Co concentration exhibited the inverted cubic structure with mixed Mn-Co sites. In agreement with a subsequent theoretical treatment, the magnetic moments of the tetragonal phases decrease with increasing cobalt content. Surprisingly, an approximately linear relationship between the number of valence electrons and the magnetic moment was found for these and other tetragonal Heusler phases~\cite{WCK+2015}, such as Mn$_2$CrGa, Mn$_3$Ga, Mn$_2$FeGa, Mn$_2$CoGa. These phases were modeled by density functional theory and a local energy minimum in the energy landscape could be observed. Nevertheless, the tetragonal phase does not show half-metallic behavior, which renders this observation particularly surprising. But this theoretical study provided evidence for the correlation of the tetragonal distortion with the peak and valley structure of the density of states.
For the case that the Fermi edge resides at a peak in the DOS, a distorted structure is likely to occur (Mn-Ga, Mn-Ni-Ga, Mn-Fe-Ga) as a slight distortion breaks either the bands' degeneracy or changes the band dispersion such that the electronic susceptibility is diminished. Mn$_2$CoGa is found to be cubic as the Fermi edge is located at a local minimum in the DOS of both spin-channels. Studying the transport properties of the Mn-Co-Ga series revealed another effect that is termed \textit{spin-selective electron localization}~\cite{WFC+2015,CDW+2015}. Spin-selective electron localization describes the change in conductivity polarization for transition metal substituted Mn$_3$Ga-systems. The exchange of Mn with a transition metal ($Y=\rm ~Sc,~Ti,~V,~Cr,~Fe,~Co,~Ni$) leads to an increase in spin-polarization since only one spin-channel is perturbed by the substitution.

\subsection{Tetragonal Compensated Ferrimagnetic Heusler Compounds{\label{sec:tetra-cchmf}}}
\begin{figure}[ht]
	\includegraphics[width=1\textwidth]{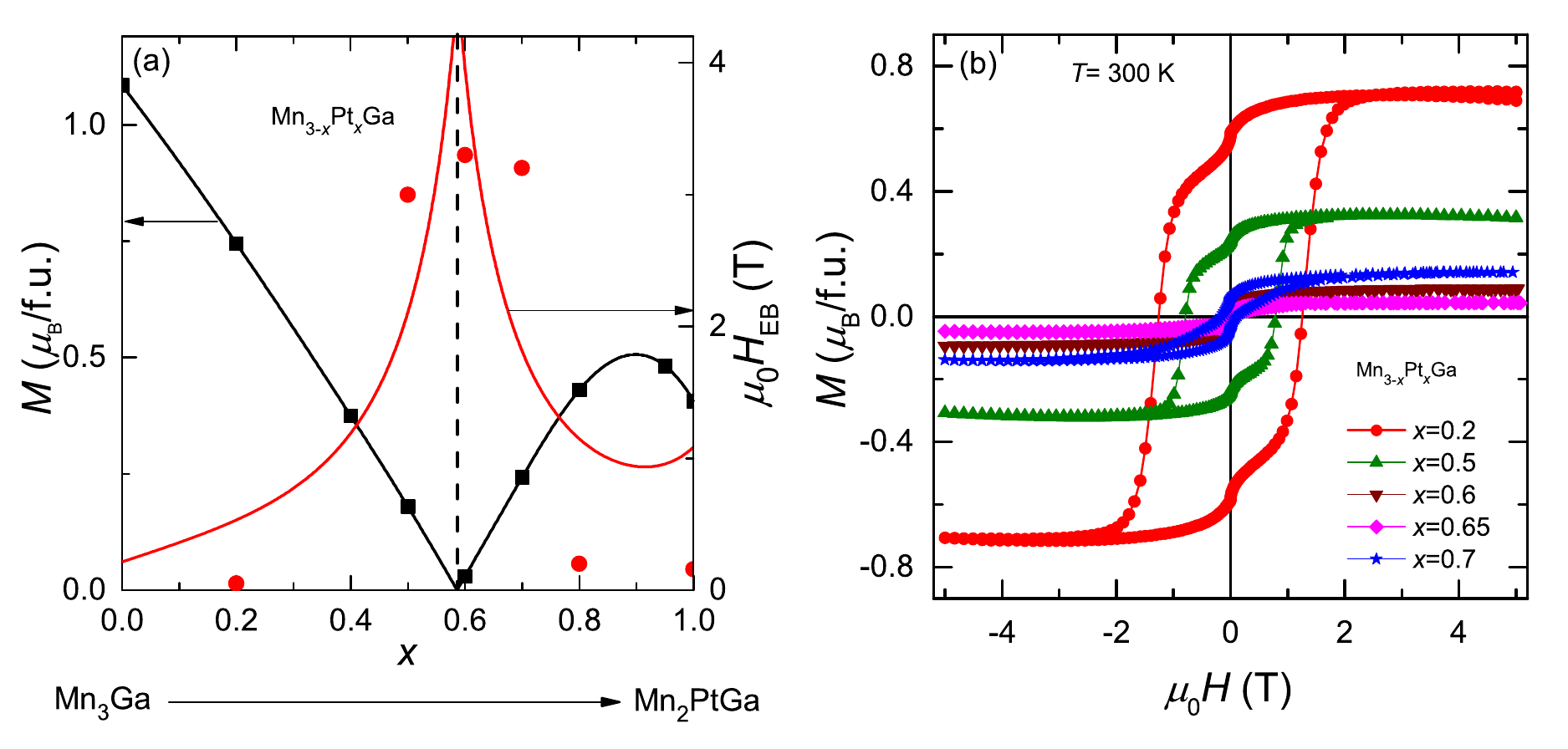}
	\caption{\label{fig4-2}(Color online)~(\textit{a})~Theoretical calculation of magnetic moment (solid squares, left axis) as a function of Pt content in Mn$_{3-x}$Pt$_x$Ga. The line connecting the solid squares is a guide to the eye.  Exchange bias field ($H_{\rm EB}$) as a function of Pt concentration $x_{\rm Pt}$ (right axis). Solid circles show the experimental data, whereas, the solid curve corresponds to a model calculation.~(\textit{b})~Curves for Mn$_{3-x}$Pt$_x$Ga thin films measured at $300\,\rm  K$.}
\end{figure}

Tetragonal magnetic materials with out of plane magnetic anisotropy are of great interest for spin transfer torque (STT) and permanent magnet related applications. It is well known that the Mn-based tetragonal Heuslers are ferrimagnetic in nature with at least two magnetic sub-lattices, where the magnetic moments align antiparallel to each other~\cite{BFW+2007,WBF+2008,WFB+2011,KJA+2011,NSW+2012,GNW+2013,KBR+2012,WCK+2014,WCK+2015,JFF+2016}. In addition,  they exhibit high magnetic ordering temperatures  well above room temperature, which is a necessary condition for any practical application. The best example of a tetragonal magnetic Heusler is Mn$_3$Ga which shows a ferrimagnetic ordering with $T_{\rm C}$ of about $750\,\rm K$~\cite{BFW+2007}. Despite these ideal characteristics, ferro/ferri magnets exhibit large unwanted stray fields that affect the magnetic state of neighboring layers in multilayer spintronic devices or neighboring devices in arrays of devices. We have proposed that a fully compensated ferrimagnetic state with a zero net magnetic moment can be achieved by systematically tuning the sub-lattice magnetic moments in Mn$_3$Ga~\cite{NNC+2015}. From theoretical calculations we have shown that the compensated magnetic state can be achieved for a wide range of Heusler materials by substituting a late transition metal element in Mn$_{3-x}Y_{x}$Ga including Ni, Cu, Rh, Pd, Ag, Ir, Pt and Au~\cite{SWS+2016}.

A specific example of magnetic compensation when $Y =\rm Pt$ is shown in Figure\,\ref{fig4-2}\,(\textit{a}).  Mn$_3$Ga consists of two non-equivalent types of Mn atoms, one within the Mn-Ga and one within the Mn-Mn planes of the tetragonal structure with space group $I4/mmm$. The Mn sitting in the Mn-Ga planes possess a larger localized moment of about $3.1\,\mu_{\rm B}$ compared to a moment of $2.1\,\mu_{\rm B}$  for the Mn sitting in the Mn-Mn- plane. However, a larger total moment of $4.2\,\mu_{\rm B}/{\rm f.u.}$ is expected from the Mn sitting in the Mn-Mn plane as there are two Mn/f.u. (formula unit) in this plane. One way of reducing the magnetic moments of the Mn sitting in the Mn-Mn plane to match that of the moment of the single Mn in the Mn-Ga plane is by partially substituting Mn in the Mn-Mn plane by a non-magnetic element. We have shown that for a Pt concentration of approximately $x=0.6$ in Mn$_{3-x}$Pt$_x$Ga, the Mn atoms in the Mn-Mn(Pt) and the Mn atom in the Mn-Ga planes contribute nearly the same moment with opposite alignment. As demonstrated in Figure\,\ref{fig4-2}\,(\textit{a}), with increasing Pt concentration, the total magnetic moment decreases and becomes zero at $x=0.6$ before further increasing for higher Pt concentrations. The theoretical concept is nicely verified by growing thin films with different Pt concentrations. As shown in Figure\,\ref{fig4-2}\,(\textit{b}), $M\mathopen{}\left(H\right)\mathclose{}$ loops measured at $300\,\rm K$ for different thin films exhibit typical out of plane hysteretic behavior found in a tetragonal system. The sample with a Pt concentration $x=0.2$, exhibits a magnetic moment of $0.7\,\mu_{\rm B}/{\rm f.u.}$ at $300\,\rm K$. The magnetization starts decreasing with increasing Pt content and becomes nearly zero for $x=0.65$ , demonstrating a fully compensated ferrimagnetic state for Mn$_{2.35}$Pt$_{0.65}$Ga. As expected, the magnetization starts increasing again for further increases in the Pt concentration.  Most importantly, these thin films show magnetic ordering temperatures well above room temperature. The feasibility of the practical application of these tetragonally compensated ferrimagnets has been demonstrated by showing the existence of a large exchange bias both in bulk and bilayer thin film materials~\cite{NNC+2013,NNC+2015,SWS+2016}.
\subsection{Non-collinear magnetic structure{\label{subsec:ncolmag}}}
\begin{figure}[ht]
	\includegraphics[width=1\textwidth]{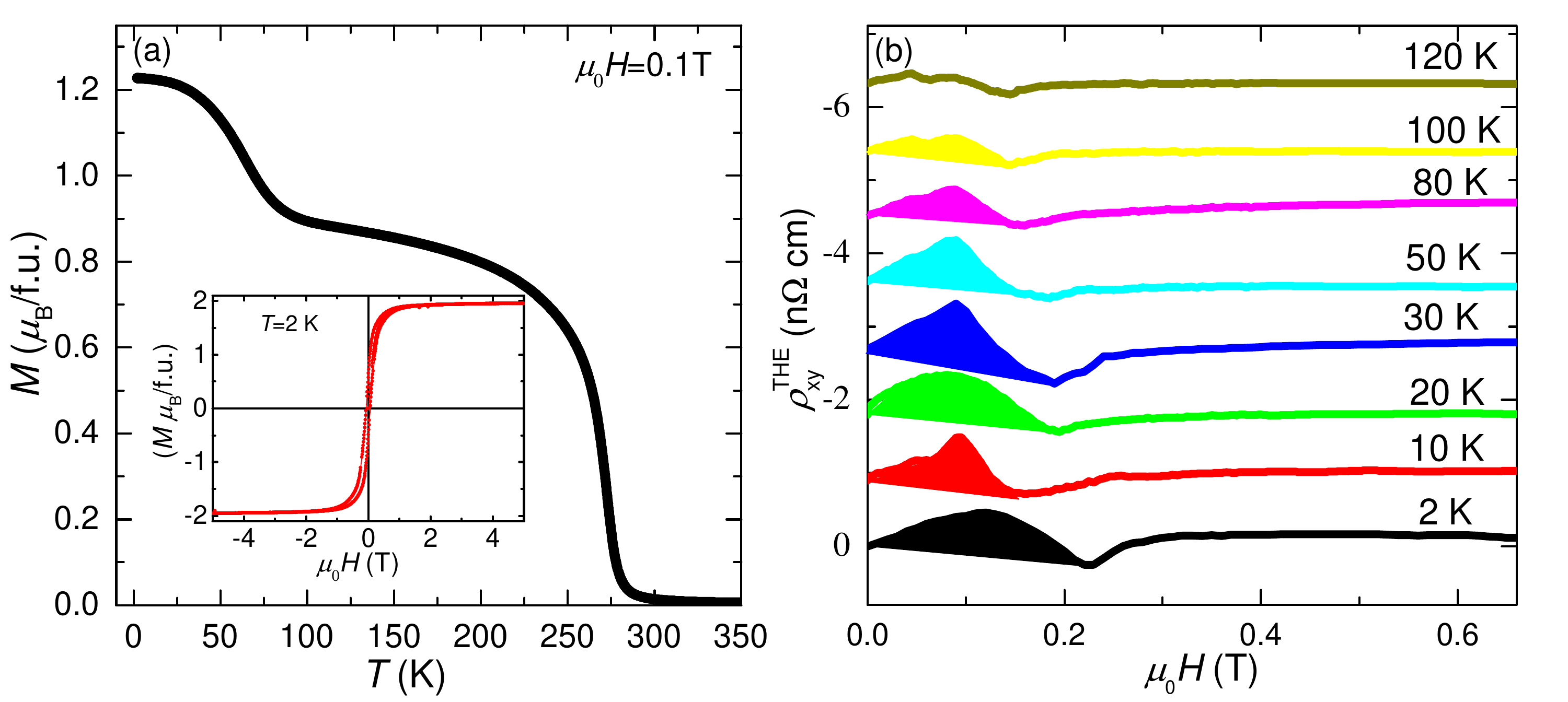}
	\caption{\label{fig4-3}(Color online)~(\textit{a})~$M\left(T\right)$ curve measured in a field of $0.1\,\rm T$ for the Heusler tetragonal compound Mn$_2$RhSn. The inset shows $M\mathopen{}\left(H\right)\mathclose{}$ loop measured at $2\,\rm K$.~(\textit{b})~Topological Hall effect ($\rho^T_{xy}$) at different temperatures for the Mn$_2$RhSn thin film.}
\end{figure}
There has been much recent interest in magnetic materials exhibiting non-collinear spin structures. One of the most exciting uses of non-collinear spin structure is the motion of chiral domain walls using spin polarized currents that generate a large chiral spin-orbit torques to drive the domain walls~\cite{RTY+2013,YRP2015}. This current driven back and forth motion of the domain walls inside a magnetic nano-wire forms a novel high density, high performance, solid-state storage memory device - Racetrack Memory - that was first proposed by Parkin \etal~in 2002 and which has the potential to even replace conventional magnetic data storage~\cite{PHT2008}. The domain wall can be replaced by other non-collinear spin textures such as skyrmions~\cite{SRB+2012}.
In this regard, Heusler materials are perfect candidates to modify the magnetic state via competing exchange interactions between different sub-lattices~\cite{NNC+2015}. In addition, most of the Mn-based Heusler materials exhibit a non-centrosymmetric crystal structure, which is necessary for the realization of the Dzyaloshinskii-Moriya (DM) interaction that leads to the formation of skyrmions~\cite{MCN+2014}.\\
The direct evidence of non-collinear magnetic state was found in the non-centrosymmetric tetragonal Heusler compound Mn2RhSn~\cite{MCN+2014}. As shown in Figure\,\ref{fig4-3}\,(\textit{a}), Mn$_2$RhSn exhibits a $T_{\rm C}$ of about $275\,\rm K$ with a transition to a state with a comparatively higher magnetic moment at about $80\,\rm K$. The magnetic ordering at the low temperature has been assigned to a spin-reorientation transition. The $M\mathopen{}\mathopen{}\left(H\right)\mathclose{}\mathclose{}$ loop measured at $2\,\rm K$ shows a saturation magnetization of about $2\,\mu_{\rm B}$ (inset of Figure\,\ref{fig4-3}), which cannot be accounted for by considering a collinear spin arrangement. In Mn$_2$RhSn, Mn atoms occupy two distinct sub-lattices. The Mn sitting in the Mn-Sn planes (MnI) show a moment of nearly $3.6\,\mu_{\rm B}$, whereas, the Mn in the Mn-Rh planes (MnII) exhibits a moment of about $3\,\mu_{\rm B}$. A simple antiparallel alignment between the moments of MnI and MnII will give a net moment of $0.6\,\mu_{\rm B}$. However, this simplified model does not match the experimentally determined magnetic moment of $2\,\mu_{\rm B}$. Similarly, a ferromagnetic ordering will result in a total moment of $6.6\,\mu_{\rm B}$, that also does not fit to the observed moment. However, the experimental magnetic moment can be explained nicely by considering a non-collinear magnetic structure in Mn$_2$RhSn. The theoretical calculations show that the competition between the antiferromagnetic interaction between the Mn moments in nearest and next-nearest planes can gives rise to a canting of the MnII moment of about 55$\degree$, thereby, providing an additional z-component of the moment to the simple ferrimagnetic configuration. The theoretical results are supported by evidence of a non-collinear spin configuration from neutron diffraction measurements~\cite{MCN+2014}.\\
Theoretical calculations also show that the present class of materials with accentric crystal structures should give rise to the formation of skyrmions under appropriate conditions. However, the experimental finding of skyrmions in bulk materials is still awaited. Recently, it has been shown that the thin films of Mn$_2$RhSn exhibit a considerable topological Hall effect (THE) as shown in Figure\,\ref{fig4-3}\,(\textit{b}). The THE was obtained in temperatures of up to $100\,\rm K$, which is the spin-reorientation transition in thin film samples. Note that materials that have skyrmions also exhibit a large THE due to their chiral non-collinear spin structures~\cite{NPB+2009,KOA+2011}. Hence, it can be assumed that there is some type of non-collinear spin structure like skyrmions present in Mn$_2$RhSn~\cite{RMK+2016}.

\clearpage
\section{HEUSLER 4.0 -- TOPOLOGICAL HEUSLER COMPOUNDS\label{part:heusler3}}
\begin{figure}[ht]
	\includegraphics[height=6cm]{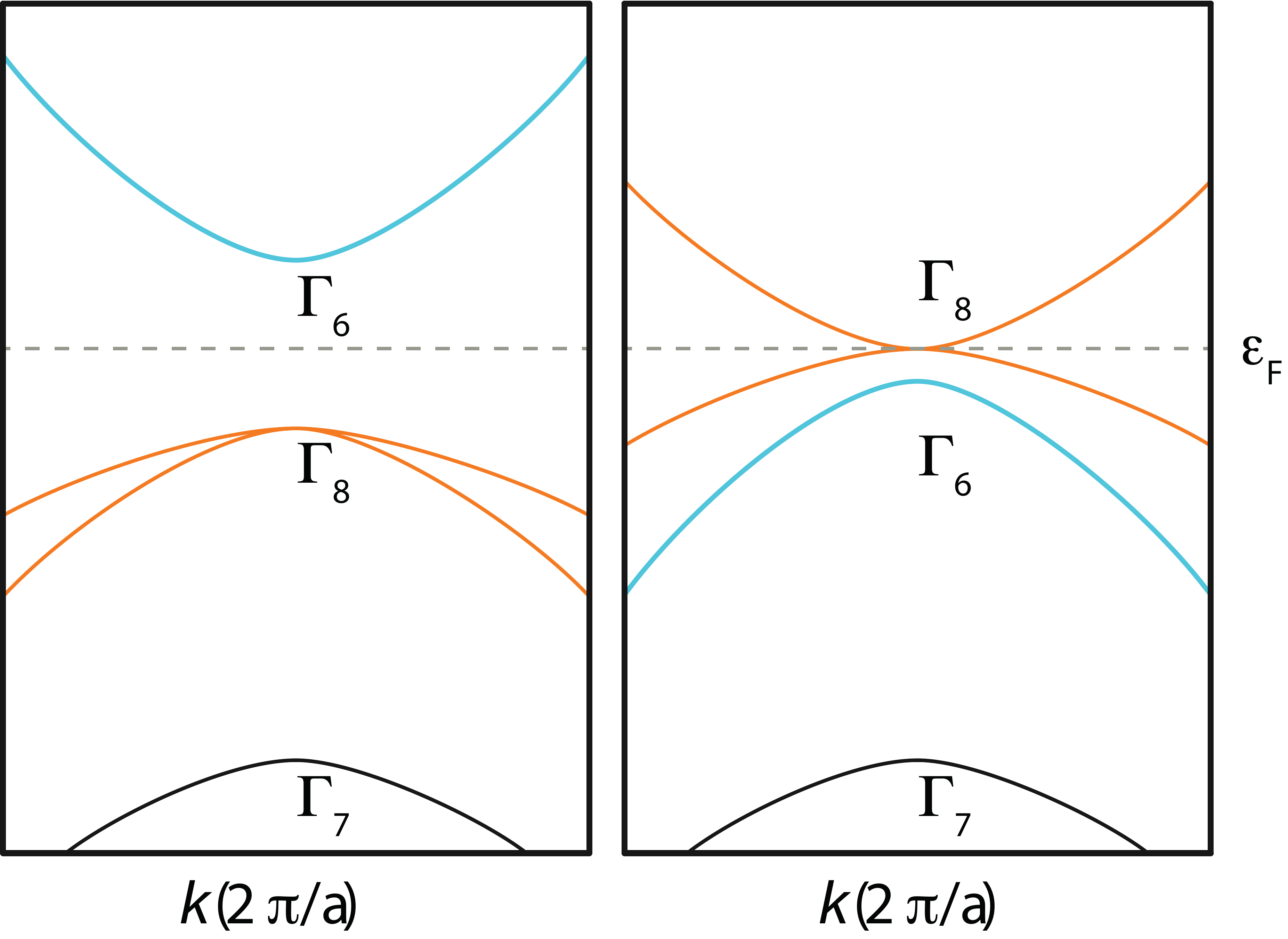}
	\caption{\label{fig4}(Color online) Schematic of a trivial and inverted band structure as found in~(\textit{a})~CdTe and~(\textit{b})~HgTe }
\end{figure}
The prediction of the Quantum Spin Hall (QSH) state~\cite{BHZ2006} triggered tremendous interest in the condensed matter community.  It not only lay the foundation for a completely new field of research, but the use and application of topological concepts in a wide range of condensed matter research gained much attention. The QSH state was subsequently realized in a HgTe/CdTe quantum well structure~\cite{KWB+2007}: materials  that crystallize in the same space group ($F\bar{4}3m$) as the $XZ$ binary semiconductors (HgTe, CdTe, ZnS) include diamond and the {\it Half}-Heusler compounds.   {\it Half}-Heusler semiconductors with 8 or 18 VEs exist with a wide range of band-gaps~\cite{BMJ+2011,OSF+2011}, and the possibility to find \textit{half}-Heuslers with a topologically non-trivial state was quickly realized~\cite{CQK+2010,LWX+2010}.  A set of \textit{half}-Heuslers that could be classified into topologically trivial and potentially non-trivial materials was proposed. Non-trivial HgTe exhibits a band-inversion of the $\Gamma_6$ and $\Gamma_8$ states at the $\Gamma$ point. Chadov \etal~could show that zero-gap semiconductors within the \textit{half}-Heusler family exhibit similar features, and an odd number of band inversions is observed in some systems. The flexibility of the Heusler compounds with resect to site occupation provides the means of tuning these materials from trivial to non-trivial states by means of lattice parameter adjustment through substitution of isoelectronic elements or adequate hybridization strength. Since the band inversion results from spin-orbit coupling, another recipe is by strengthening of the average spin-orbit interaction via, for example, substitution by heavy transition metals. Some of the zero-gap semiconductors that have been experimentally realized show a surprising range of multifunctional properties in addition to the non-trivial electronic structure, when rare earth elements are introduced, as for example: GdPtBi -- antiferromagnetism~\cite{CTB+1991}, LaPtBi -- superconductivity ($T_{\rm crit.}< 0.9\,\rm K$)~\cite{GMH+2008}, YPtBi -- heavy fermion behavior and superconductivity ($T_{\rm crit.}< 0.9\,\rm K$)~\cite{FCB+1991,BSK+2011}, ErPtBi -- antiferromagnetism ($T_{\rm N}< 1.2\,\rm K$)~\cite{CTB+1991}. The zero-gap state, that is a prerequisite for topological insulators, is also a useful feature for thermoelectric materials. Consequently, a connection between topological insulators, the zero gap-state, and thermoelectric performance is profound~\cite{MCY+2013}.

\subsection{GdPtBi -- a Weyl-Semimetal in a magnetic field{\label{sec:gdptbi}}}
\begin{figure}[ht]
	\includegraphics[width=\textwidth]{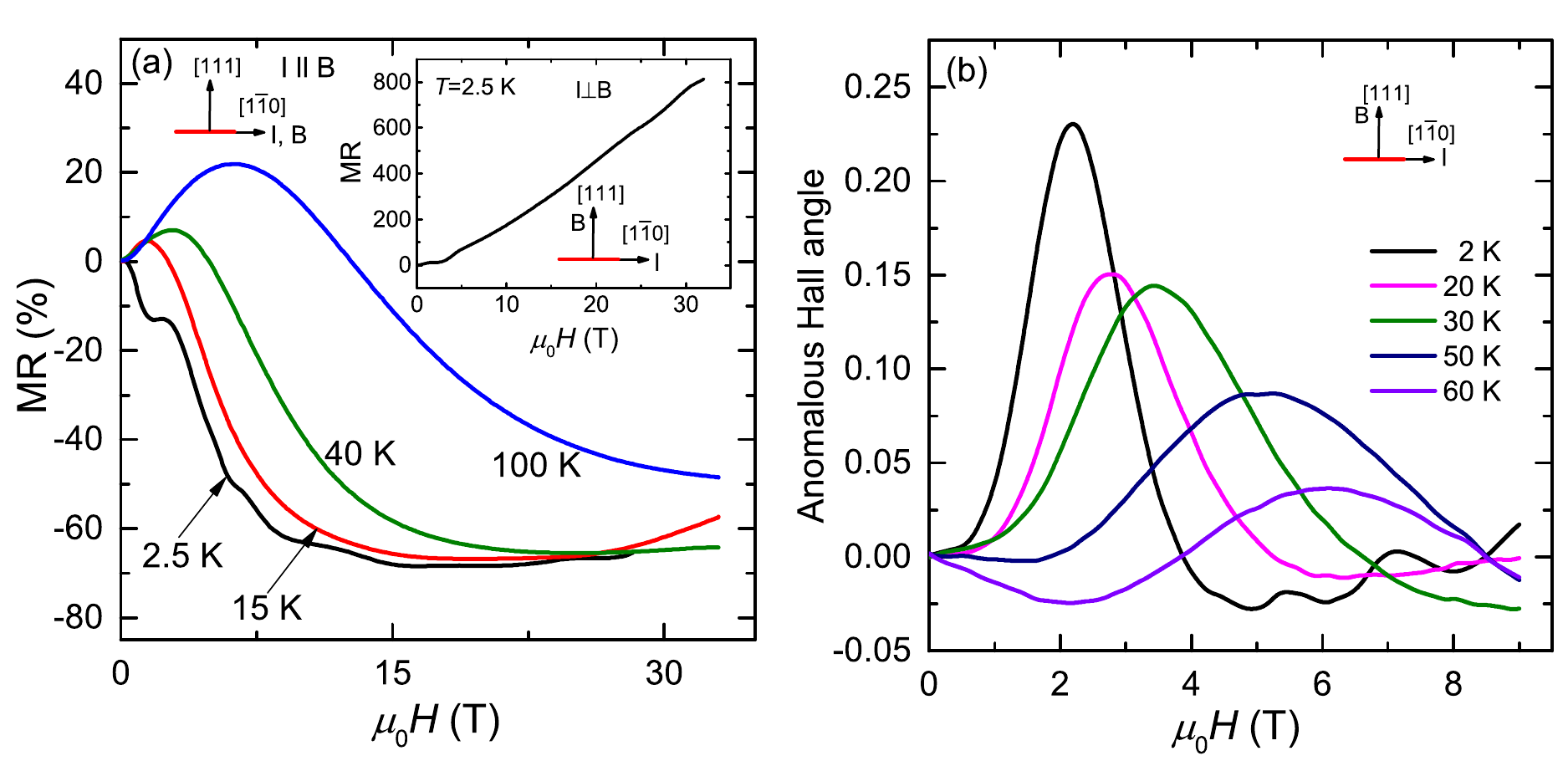}
	\caption{\label{fig3-2}(Color online)~(\textit{a})~Field dependence of longitudinal magnetoresistance (MR) measured at different temperatures for fields up to $33\,\rm T$ in GdPtBi for the crystal directions shown in the figure. The inset shows the transverse MR measured at $2.5\,\rm K$.~(\textit{b})~Anomalous Hall angle (AHA) at different temperatures for GdPtBi.}
\end{figure}
Weyl semimetals (WSM) are a class of topological semimetals, beyond topological insulators, where the conduction and valence bands cross in the vicinity of the Fermi level~\cite{WFF+2015,SNS+2015,HZL+2015,Ciudad2015,XBA+2015,HXB+2015,XAB+2015,XBS+2015,XKL+2015,Dai2015,LYS+2015,BMA+2016,CXS+2016}. The crossing points are called Weyl points that are separated in momentum space. These Weyl points, which act as magnetic monopoles in momentum space, always appear in pairs and are connected by an unusual surface state termed the Fermi arc~\cite{Dai2015,LYS+2015,LXH+2015,CXZ+2016,BXS+2016,BMA+2016}. In general, semimetals exhibit ultra-high carrier mobilities and large transverse magnetoresistance values (MR), that are also promising effects for spintronic applications~\cite{Yang1999,XHR+1997,AXF+2014}. However, the non-trivial electronic structure of a WSM can give rise to additional unusual phenomena such as negative magnetoresistance which is connected to chiral magnetic effects~\cite{HZL+2015,XKL+2015}. The chiral anomaly comes from the charge pumping between two Weyl points connected through the Fermi arc. In most of the three dimensional WSMs, where the breaking of either time reversal symmetry or inversion symmetry occurs, the Weyl points come as a result of accidental touching/crossing of the conduction and valence bands. In recent work, however, experimental evidence of an unusual topological surface state~\cite{LYW+2016a} and a chiral anomaly~\cite{HKW+2016} have been found in the lanthanide \textit{half}-Heusler semimetals.\\
The existence of negative MR that is associated with a chiral anomaly in these lanthanide \textit{half}-Heuslers was attributed to the formation of Weyl points due to an external field induced Zeeman splitting. However, it is presumed that the Zeeman splitting is negligible in comparison to the much larger exchange-field coming from the 4\textit{f} electrons in \textit{RE}PtBi compounds ($RE=\rm~Gd,~ Nd$).  Both GdPtBi and NdPtBi are antiferromagnetic below their respective N\'eel temperatures of $9.0\,\rm K$ and $2.1\,\rm K$~\cite{KKK+2011,MLL+2014,MDG+2015}. In the absence of an external field, the exchange fields originating from the magnetic moments at different sub-lattices cancel and, hence, no Weyl points are observed. With the application of a modest external magnetic field the Gd moments in different sub-lattices align parallel to each other resulting in a ferromagnetic ordering. Thus, the exchange-splitting of the conduction bands can give rise to the formation of Weyl points.\\
The existence of a large unsaturated longitudinal negative MR in GdPtBi is shown in Figure\,\ref{fig3-2}\,(\textit{a}). The MR measurements have been performed up to fields of $33\,\rm T$ with $I$ parallel to $B$ with the configuration shown in the figure. A large negative MR of $~68\%$ is observed at $2.5\,\rm K$. The negative MR persists up to temperatures greater than $100\,\rm K$, which is much above the $T_{\rm N}$ of GdPtBi~\cite{SNS+2016}. However, the field required to give a negative MR increases with increasing temperature due to thermal effects. As shown in the inset of Figure\,\ref{fig3-2}\,(\textit{a}), a large transverse positive MR is observed when the magnetic field is applied perpendicular to the current. These experiments confirm that the negative MR in GdPtBi originates from the formation of Weyl points due to the field driven splitting of the conduction band. Another proof of the chiral anomaly is the finding of a large anomalous Hall effect in GdPtBi. A large anomalous Hall angle (AHA) of about 0.23 is calculated in a field of $2.2\rm\, T$ at $2\rm\, K$. The AHA decreases with increasing temperature and the field where a maximum AHE is observed shifts to higher magnetic fields. This indicate that the AHE is closely related to the presence of negative MR originating from the chiral anomaly. 

\subsection{Co$_2$TiSn -- a magnetic, centrosymmetric Weyl semimetal{\label{sec:co2tisn}}}
\begin{figure}[ht]
	\includegraphics[width=\textwidth]{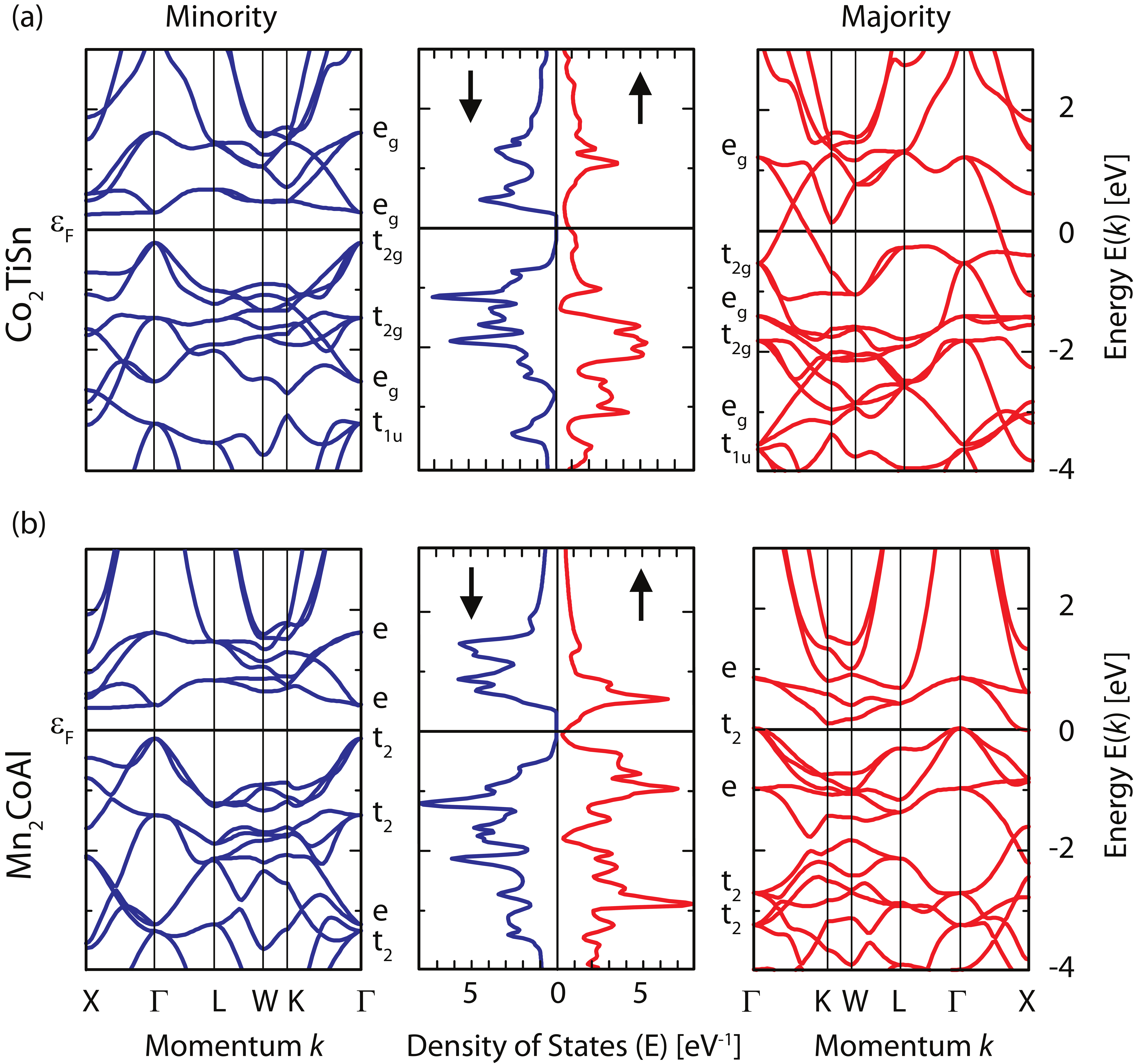}
	\caption{\label{fig3-1}(Color online) Spin resolved band structure and density of states of~(\textit{a})~the magnetic Weyl semimetal Co$_2$TiSn and~(\textit{b})~the spin-gapless semiconductor Mn$_2$CoAl, both with 26 valence electrons.}
\end{figure}

Crossing points in the electronic band structure, such as Weyl and Dirac points, are not rare, but very often these points are not at the Fermi energy or many other bands also cross the Fermi energy. Weyl and Dirac points have a strong influence on the transport properties and other properties if they are very close to the Fermi energy in a semimetal. It is established and nicely described in the review of Nagaosa~\etal~\cite{NSO+2010} that these crossing points in the electronic structure lead to an enhanced Berry phase and, therefore, to a large anomalous Hall effect (AHE) in magnetic systems  and  to a spin Hall effect (SHE) in paramagnetic metals or diamagnetic topological semimetals. It is reported that these points act as a magnetic monopole for the Berry curvature in momentum space~\cite{FNT+2003}. A large AHE in connection with a large Berry phase in Heusler compounds was predicted by K{\"u}bler and Felser for Co$_2$MnAl~\cite{KF2012}. The predicted value of the giant AHE was in excellent agreement with the experimentally determined value~\cite{VSS+2011}.  Husmann~\etal~\cite{HS2006} investigated Co$_2$CrAl and found that the intrinsic AHE dominates. An intrinsic contribution comes from the Berry phase while extrinsic contributions originate from scattering from impurities through skew scattering and side jump contributions. It is noteworthy that K{\"u}bler and Felser had already noted a Dirac point below the Fermi energy in Co$_2$VSn, but a direct connection between the giant AHC and the Dirac point was not made~\cite{KF2012}. In the context of recent searches for Weyl and Dirac points in the electronic structure of semimetals Wang\etal~\cite{WVK+2016} and Chang~\etal~\cite{CXZ+2016c} proposed that electron doped Co$_2$TiSn should be a magnetic Weyl semimetal. Based on the work on AHE, K{\"u}bler and Felser recognized that Co$_2$MnAl is a Weyl metal even when undoped. It seems that there is a relation between the Weyl semimetals and the spin gapless semimetals. SGSs appear only in the non-centrosymmetric space group 216 ($F\bar{4}3m$), whereas for the same number of valence electrons the corresponding Heusler compounds in space group 225 ($Fm\bar{3}m$) are Weyl semimetals. The minority band structures for the respective materials are identical in both space groups, but the bands are not allowed to cross in the space group 225 and turn to a forbidden crossing and therefore a semiconducting majority spin-channel is obtained instead of a Weyl semimetal (See Figure~\ref{fig3-1}). This concept can help to identify Weyl semiconductors and non centro-symmetric semiconductors via simple electron counting rules.

\subsubsection{Topological non-cubic Heuslers{\label{subsec:topononcub}}}
\begin{figure}[ht]
	\includegraphics[height=8cm]{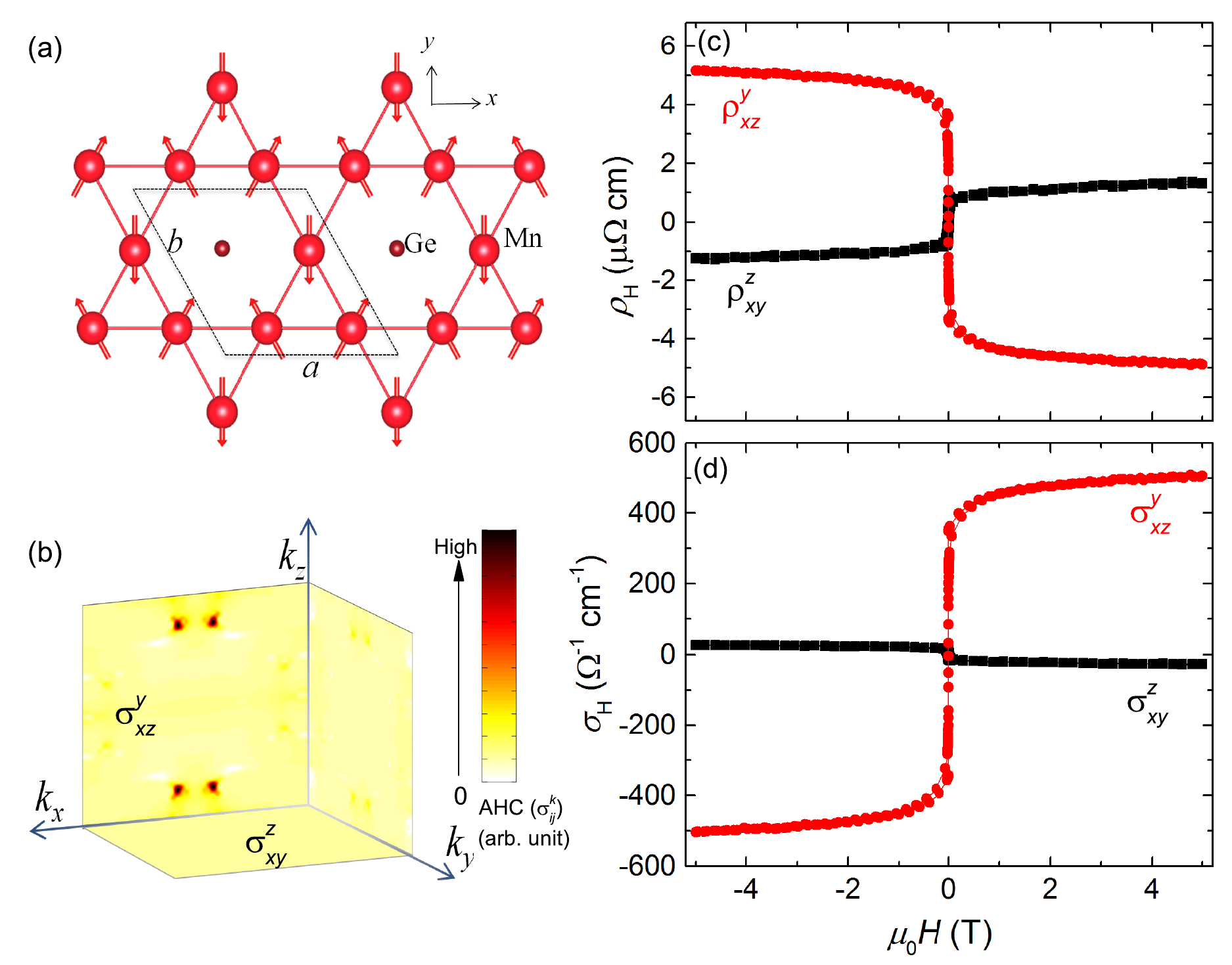}
	\caption{\label{fig4-4}(Color online)~(\textit{a})~Triangular non-collinear antiferromagnetic configuration for hexagonal Mn$_3$Ge.~(\textit{b})~Berry curvature integrated along the $k_x$, $k_y$ and $k_z$ axes, plotted in the $k_y–k_z$, $k_x–k_z$ and $k_x–k_y$ planes, respectively. For the spin configuration shown in a, only a nonzero anomalous Hall conductivity ($\sigma^y_{xz}$) is obtained in the $k_x-k_z$ plane.~(\textit{c})~Hall resistivity ($\rho_{\rm H}$) as a function of magnetic field ($H$) measured at $2\,\rm K$.~(\textit{d})~Field dependence of Hall conductivity ($\sigma_{\rm H}$) obtained from the corresponding Hall resistivities shown in (\textit{c}).}
\end{figure}
The existence of a non-collinear magnetic structure in tetragonal Heusler materials is discussed in the previous section~\cite{MCN+2014,RMK+2016}. Many of these Mn-based Heusler materials also exhibit a stable hexagonal crystal structure~\cite{NTY1982,TYN1983,YSM+1988,QNK+2014}. By varying the preparation conditions cubic, tetragonal and hexagonal phase can be stabilized in one system~\cite{ZYW+2013}. It turns out that most of the hexagonal materials display an antiferromagnetic ordering~\cite{NTY1982}. Neutron diffraction studies of the hexagonal Mn$_3$Sn and Mn$_3$Ge compounds reveal the presence of a non-collinear antiferromagnetic ordering~\cite{YSM+1988}. In particular, hexagonal Mn$_3$Ge, which consists of two layers of Mn triangles stacked along the c-axis, shows a 120$\degree$-triangular antiferromagnetic structure~\cite{YSM+1988}. As shown in Figure\,\ref{fig4-4}\,(\textit{a}), the Mn atoms form a Kagome lattice with Ge sitting at the center of a hexagon. The non-collinear spin structure arises due from geometrical frustration of the Mn spins arranged in the triangular spin structure. Recent theoretical works have demonstrated that materials with non-collinear antiferromagnetic structures and with some special symmetries should exhibit a large anomalous Halle effect (AHE)~\cite{CNM2014,KF2014}. Note that the AHE is an intrinsic property of all ferromagnets and roughly scales with the magnetization~\cite{NSO+2010}. Therefore, antiferromagnets with a zero net magnetic moment, in general, should not display an AHE. Since the intrinsic AHE is a manifestation of the Berry curvature, we have calculated the Berry curvature in Mn$_3$Ge by considering the non-collinear antiferromagnetic spin structure shown in Figure\,\ref{fig4-4}\,(\textit{a}). The distribution of Berry curvature in the momentum space is depicted in Figure\,\ref{fig4-4}\,(\textit{b}). It can be clearly seen that a non-vanishing Berry curvature is only obtained in the $k_x-k_z$ plane, and almost zero amplitudes of the Berry curvature are observed in other two planes.~\cite{NFS+2016} 
Our theoretical predictions were supported by measuring the experimental AHE in Mn$_3$Ge, as shown in Figure\,\ref{fig4-4}\,(\textit{c}) and (\textit{d}). A large anomalous Hall resistivity ($\rho_{\rm H}$) of $5.1\,\mu\Omega$ cm was found in the $x-z$ plane ($\rho_{xz}^y$), whereas, a small ($\rho_{\rm H}$) of about  $5.1\,\mu\Omega$ was measured in the $x-y$ plane ($\rho_{xy}^z$). Similarly, an extremely large AHC of about $500 \Omega{\rm cm}^-1$ and almost zero AHC are calculated in the $x-z$ ($\sigma_{xz}^y$)  and $x-y$ ($\sigma_{xy}^z$)  planes, respectively~\cite{NFS+2016}. A similarly large AHE has been also been found in the hexagonal non-collinear antiferromagnetic Mn$_3$Sn~\cite{Nakatsuji_2015}. It is worth noting that the ferromagnetic Heusler compounds Co$_2$MnSi and Co$_2$MnGe with a net magnetic moment of about $5\,\mu_{\rm B}$ display a maximum $\rho_{\rm H}$ of $4\,\mu\Omega$~\cite{WFK+2005}.  Thus, the finding of such a large AHE in the antiferromagnetic Mn$_3$Ge and Mn$_3$Sn can be explored further for their possible use in antiferromagnetic spintronics~\cite{JMW+2016}.

\section{SUMMARY \label{part:summary}}
\begin{summary}

The Heusler compounds have a rich history dating back more than a century and yet fascinating new properties continue to emerge even today.  In this review we have elaborated the history of Heuslers in 4 distinct development phases. In Heusler 1.0 the discovery of combinations of essentially non-magnetic elements that form ferromagnetic compounds well above room temperature was a remarkable finding.  In Heusler 2.0 the theoretical prediction and experimental finding of half-metallicity in certain classes of Heusler materials was another major discovery that would lead several decades later to the funding of giant value of tunneling magnetoresistance. In Heusler 4.0 more complex magnetic structures in which the magnetic moments are aligned non-collinearly have been discovered in a range of Heusler compounds. In Heusler 3.0 an entirely new world of Heusler compounds was revealed by the application of newly developed notions of topology that led to the prediction and later experimental proof of topological insulators and, more recently, Weyl semi-metallic Heuslers. What is perhaps even more remarkable is that many of these properties evolve from simple concepts of electron counting. By changing the number of valence electrons the magnetization and Curie temperature of magnetic Heuslers can be varied systematically or the Weyl points in their band structure can be tuned though the Fermi energy. In this review there was no space to describe other properties of the Heuslers that include, for example, giant thermoelectricity or the magnetocaloric applications. Heusler compounds are promising for a wide range of applications that has focussed to date on spintronic applications.  The future for Heuslers seems very bright.

\end{summary}

\section*{DISCLOSURE STATEMENT}
The authors are not aware of any affiliations, memberships, funding, or financial holdings that
might be perceived as affecting the objectivity of this review. 

\section*{ACKNOWLEDGMENTS}
Financial support from the \textsl{European Research Council Advanced Grant} (ERC-AG; No.~291472~``IDEA~Heusler!'') is gratefully acknowledged.

\end{document}